\documentstyle[twocolumn,aps]{revtex}
\begin{document}
\draft
\title{\bf {Features of level broadening in a ring-stub system}}
\author{P. Singha Deo$^{a,}$ \cite{eml} and M. V. Moskalets$^b$}
\address{$^a$Department of Physics, University of Jyv{\"a}skyl{\"a},
      P.O.Box-35, 40351 Jyv{\"a}skyl{\"a}, Finland\\
      $^b$Fl.48, 93-a prospect Il'icha, 310020 Khar'kov, Ukraine}
\maketitle
\begin{abstract}
When a one dimensional (1D) ring-stub system 
is coupled to an electron reservoir, the states acquire a
width (or broadening characterized by poles in the complex
energy plane) due to finite life time effects. We show that this
broadening is limited by anti-resonances due to the
stub. The differences in level broadening in presence and absence
of anti-resonance is exemplified by comparison to a 1D
ring coupled to an infinite reservoir. We also show that the
anti-resonances due to the stub has an anchoring effect on
the poles when a magnetic flux through the ring
is varied. This will have
implication on change in distribution of the poles in disordered
multichannel situation as magnetic flux is varied.
\end{abstract}
\pacs{PACS numbers: 72.10.-d; 73.20.Dx}
\narrowtext

Dephasing in mesoscopic systems is of current research interest
and there are various models for dephasing \cite{par} in
situations where there is a spatial separation between elastic
and  inelastic processes. As such it is a difficult problem to
model dissipation theoretically because dissipative systems are
open systems which are in general more difficult to deal with
than closed systems. Real systems are on the other hand most of
the time dissipative and it is extremely difficult to obtain a
non-dissipative sample in the laboratory. A pioneering idea of
Landauer offered a very simple approach to dissipation in
electronic systems when he showed that an electron reservoir can
introduce some basic features of dissipation like time
irreversibility and resistance \cite{lan}. This idea evoked a
lot of research and resulted in the Landauer-B{\"u}ttiker 
conductance formula which is now experimentally realized to be
correct \cite{but1}. The idea was later extended by B{\"u}ttiker
\cite{but2} to introduce level broadening in a ring (or
any finite system) penetrated
by magnetic flux and there is at least one experiment which
verifies the magnitude of persistent current in such a open
system \cite{mai}. Some more precise testing ground for
persistent currents in open systems could be the experimental
observation of its directional dependence i.e., if a persistent
current loop is coupled to a DC current carrying quantum wire
then the magnitude of persistent current will depend on the
direction of the DC current \cite{jay1}. It is to be noted that
noise or the DC current itself does not depend on
directions. It was also shown that a quantum
current magnification effect is possible with a DC bias voltage
\cite{jay2}
or with a temperature difference \cite{mos}
within this theoretical frame work when one can have large
circulating currents in the ring in the absence of an
Aharonov-Bohm flux.

In the present work, we apply this model for level broadening to
study a simple system that consists of a one dimensional (1D)
ring coupled to a 1D side branch (or stub). 
This elementary system has received some attention in
the past where it was shown that one can have persistent
currents without parity effect \cite{deo1}, coulomb blockade
\cite{but3},  bistability \cite{and}, etc. Parity effect of
persistent currents in single channel rings mean consecutive
states have opposite slopes \cite{leg},  whereas in the
ring-stub system there are two kinds of eigenstates: parity
conserving and parity violating \cite{deo1}. There are $l/u$
number of consecutive states forming a group with the same slope
after which there are $l/u$ consecutive states with slope
opposite to the those in the first group \cite{deo1}. Here $l$
is the length of the stub and $u$ is the length of the ring. Two
consecutive states with same slope are called parity violating
pair whereas two consecutive states with opposite slope are
called parity conserving pair. 

When the side branch is adjusted one can have situations when
the length of the ring is smaller than the phase coherence
length $l_\phi$ while the length of the side branch is smaller,
comparable or greater than $l_\phi$. When the side branch is
much larger than $l_\phi$ then the situation is the same
as a ring coupled to a reservoir.
One can also apply the known mechanism of level
broadening of a ring-stub system
by coupling to a reservoir which is completely
physical as long as the stub length is smaller than $l_\phi$ but
can be much larger than the ring length. We show that 
in this regime the
presence of anti-resonances that always occur
in quantum wires \cite{deo3} has drastic effects on
level broadening. Only
in the absence of such anti-resonances
the persistent current in a ring coupled to a large stub with
some level broadening is similar to persistent current in a ring
coupled to a reservoir. These anti-resonances
will also affect
the transition between different universality classes in Random
Matrix Theory in disordered multichannel situations.

The ring-stub system coupled to a reservoir is shown in Fig.~1.
A flux $\phi$ penetrates the ring. An electron reservoir can be
attached in three ways as shown in (a), (b) and (c). In (b) and
(c) there are two junctions in the set up. One is the
reservoir-system junction (X) and the other is the ring-stub
junction (Y). In (a) the two junctions are merged into one and
there is only a single junction (Z).  While (a) is a ring-stub
system coupled to an electron reservoir, (b) and (c) will have
some additional features due to multiple scattering and
resonance between points X and Y. However, (b) and (c) will
behave similarly to each other which leads us to exclude the
situation in (c). In (b) the length of the ring is $u$, the
distance between X and Y is $v$ and the distance between X and
the dead end of the stub is $w$. Hence length of the stub is
$l=v+w$. $u$ is taken to be the unit of length and we will
mention the other lengths in numbers without mentioning the
units. When $v \rightarrow 0$ the the system in (b) continuously
goes over to the system in (a). Free particle quantum mechanical
wave function can be written down in different regions of the
system in (b) and can be matched at the junctions using the
Griffith  boundary conditions \cite{gri} or the three way
splitter S-matrix where $\epsilon$ 
determines the strength of coupling \cite{but2}. Analytical
expression for the persistent current $dI$ in an wave vector
interval $dk$ is given below using the S matrix approach for the
system in Fig.~1(b).
$$dI/dk=\frac{e\hbar k}{2\pi m}
   (-1) \epsilon_2 \epsilon_1\sin(\alpha) \sin(k u)\sin^2(k w)/D$$
   $$D=4 a_2^2\sin^2(k w) A^2 + b_2^2 B^2$$
   $$A=b_1 \sin(k v) [\cos(\alpha)-\cos(k u)] + a_1\sin(k u)\cos(k v)$$
   \begin{equation}
   B=b_1\sin(kl) [\cos(\alpha)-\cos(k u)] +
   a_1\sin(k u)\cos(kl).
   \end{equation}
Where
\begin{equation}
a_i={1\over 2}(\sqrt{1-2\epsilon_i}-1), \qquad 
b_i={1\over 2}(\sqrt{1-2\epsilon_i}+1)
\end{equation}
$i=1$ for the ring-stub junction and 2 for the reservoir-stub
junction. Here $\alpha = 2\pi \phi /\phi_0$, $\phi_0$ being
the flux quantum.

There can be two kinds of processes in the junctions that can
lead to reflection of an electron waveguide. First is due to
diffraction as the wave front splits up at the junction which
disappears for $\epsilon=0.5$. For smaller values of $\epsilon$
this contribution to reflection is always there. $\epsilon=4/9$
corresponds to Griffith boundary conditions exactly for a free
junction.  The second is the reflection due to the weak coupling
or a potential scatterer at the junction
that leads to weak coupling. However,
there is a third way of getting a reflection at the junction (X)
in (b) and that is due to an interference effect that
produces an anti-resonance. Such anti-resonances
occur very generally in a quantum wire of finite width
due to evanescent modes \cite{deo3} that
can be mapped exactly into the 1D stub model \cite{deo3}. 
An electron coming from
the reservoir on reaching junction X can go towards the ring or
can go towards the dead end of the stub, get reflected back and
then go towards the ring. Interference between these two paths
can be constructive or destructive depending on the wave vector
and can lead to reflection from the junction X. We are here
discussing first order reflection from the junction X which
determines the strength of coupling between the reservoir and
system. Besides this there is always second order reflection
(reflection from other junctions in the system partly flow out
of the junction X towards the reservoir) that makes the total
current in the lead to be always zero.

When the state of the
ring-stub system is broadened by the reservoir, each broadened
state will have a pole in the reflection amplitude at junction X
that behave similarly as the eigen energies as the flux is
varied and the persistent current at the broadened areas can be
described using the on shell scattering matrix \cite{akk1}.  In
Fig.~2 we plot the persistent current in an infinitesimal energy
range (${2 \pi \hbar \over e} {dI \over dE}$) 
versus incident wave vector $ku$ for this system
for two different flux values. The solid  curve is for $u$=1,
$v$=0.01, $w$=9.99, $\alpha=2\pi \phi/ \phi_0$=0.1 and for
Griffith boundary conditions at the free junctions. Here $\phi$
is the flux through the ring and $\phi_0$ is the  flux quantum.
The dotted curve is for $\alpha=1.5$ with other parameters
remaining the same. We have only plotted up to $ku$=$2\pi$
because at higher energies the curve repeats itself
qualitatively. It can be noted that in the solid curve there are
10 diamagnetic peaks ($l/u$ being 10) consecutively, followed
by 10 paramagnetic
peaks. As the flux is increased we get the dotted curve in which
the diamagnetic peaks shift to higher energy and paramagnetic
peaks shift to lower energy compared to the solid curve.
Diamagnetic states (broadened by the coupling to the
reservoir) group together because of a discontinuous phase
change as Fermi energy crosses the zero \cite{deo2} (or
anti-resonance) in the persistent current
between each broadened peak which arises because of total first
order reflection at X due to the interference effect discussed
above. Level
broadening in this case is limited by the presence of zeroes
(or anti-resonances) and
the peaks do not overlap with each other. As a result when the
stub is made very long, the persistent current in the ring
coupled to a stub does not bear any similarity with that of a
ring coupled to a reservoir. This is shown in Fig.~3 where
dotted curve is for $u$=1, $v$=0.01, $w$=99.99 and $\alpha$=1.0
for Griffith boundary conditions. The thick curve is the
persistent current in a ring of length $u$ coupled to an
infinite reservoir for $\alpha=1.0$.  The two curves have no
similarity at all.
Also this is a very simple example that show the
zeroes anchor the poles and the poles cannot move freely as the
magnetic field is varied. The magnetic field has no effect on
the zeroes because the zeroes are determined by the 
localized states of
the stub. This anchoring effect of the zeroes on the poles can
drastically change the distribution of the poles in disordered
systems like that considered in Ref. \cite{akk2} i.e., a
disordered multichannel ring threaded by a magnetic flux. The
magnetic flux is known to make the eigen energies rigid which in
turn leads to a transition between different universality
classes of level statistics in Random Matrix Theory. The
anchoring effect of the zeroes, that will be unaffected by the
magnetic flux, will substantially add color to this rigidity
phenomenon.

The effect of the anti-resonances on transport currents has
been studied to some extent \cite{bay}. Here we have shown
its effects on persistent currents. Transport currents
or transmission coefficient being independent of
magnetic flux and
bounded by unity does not exhibit the drastic effects
shown in Figs.~2 and 3. To exemplify this further let us
study a situation when there are no anti-resonances.
The features of level broadening and level statistics
when a finite size system is coupled to a reservoir is
studied in Ref. \cite{kon} when there are no anti-resonances.
So the features that we will obtain in the following are
in accordance with the theory developed in Ref. \cite{kon}
but completely different from the situation in Figs.~2 and 3.
Essentially in the following
we will get a situation that can continuously
go over to a ring-reservoir system when the stub length
becomes large.

The anti-resonances can be removed by a different 
boundary condition at the dead end of the stub \cite{pri}
instead of a hard wall boundary condition as used in this work
or by a magnetic field in the stub region if the stub is
quasi-one dimensional etc,. In order to show this we use a simple
trick to remove the first order total reflection at the junction
X. We make a special choice of parameters that is $u$=1,
$v$=9.99 and $w$=0.01 for Griffith boundary conditions. In this
case the first order total reflection occur at $ku$=0, 100$\pi$,
200$\pi$ and so on. At $ku$=50$\pi$, 150$\pi$, 250$\pi$ etc, the
first order reflection at X is zero. Hence in an energy regime
like $ku$=20$\pi$ to 22$\pi$ the first order reflection
coefficient at X is approximately 0.6 and almost independent of
energy. So in this energy window of $2\pi$ the ring-stub system
is weakly coupled to a reservoir for such parameters.  In Fig.~4
the thin solid curve is the persistent current versus $ku$ for
$\alpha$=1.5 and the thick solid curve is that for $\alpha$=0.1.
The broadening is already enough to make the resonances overlap
with each other. In the absence of the anchoring effect on the
resonances, some resonances can shift a lot with the magnetic field
as compared to that in Fig.~2. 
The dashed curve and  the dotted
curve are the persistent current in a ring coupled to an
infinite reservoir at $\alpha$=1.5 and 0.1, respectively.
Keeping all parameters same as in Fig.~4 we plot the same things
in Fig.~5 in a different energy window (40$\pi$ to 42$\pi$)
where the first order reflection at the junction X is 0.18.
Hence this is a situation of strong coupling between the system
and the reservoir and the peaks have broadened further compared
to Fig.~4.  Curve conventions are the same as in Fig.~4.  Now
keeping the ring length ($u$=1) to be the same we make the stub
very long i.e. $v$=99.99 and $w$=0.01 and plot persistent
current versus $ku$ in the same energy interval as that in
Fig.~5, for a magnetic field that gives $\alpha$=1.0 in Fig.~6.
Here we have just interchanged the values of $v$ and $w$ as
compared to that in Fig.~3 and we have exactly the same number
of poles as that in Fig.~3. Persistent current of a ring of
length $u$=1 connected to an infinite reservoir is shown by the
solid curve. 

From Figs.~4, 5 and 6 it is seen that as the length
of the stub is made longer and broadening of the states of the
ring-stub system is made larger, the persistent current
is rapidly oscillating around a mean
value which turns out to be the persistent current in a ring
coupled to an infinite reservoir. The rapid oscillations are
resonance effects that are again due to the fact that there are
no inelastic processes in the stub. A finite number of inelastic
processes (which will arise in a situation when the stub length
is comparable to $l_\phi$) will smear out these oscillations and
then the situation in Fig.~6 will be similar to that of a ring
coupled to a reservoir. But the situation in Fig.~3 is different
because of the limiting effects of the anti-resonances. 
Most of the large
current carrying peaks occur around a region where the thick
solid curve is very small. Inelastic processes are expected to
affect the different peaks equally and for a finite number of
inelastic processes a difference between the two situations in
Fig.~3 and 6 will continue to exist. However, when the stub
length becomes much larger than $l_\phi$ then there will be a
saturation of the broadening effect \cite{but2} it produces on
the resonances. Once this saturation effect sets in, the
broadening will affect the larger resonances more than the
smaller ones. Such situations cannot be studied in this model.
Finally in both cases one will get the situation when the stub
becomes a part of the reservoir and the persistent
current in the ring will be that of a ring coupled to
a reservoir. One can design similar problems with
transport current across the stub but as mentioned before,
the transmission coefficient being independent
of flux and bounded by unity,
the features are not so prominent.
We hope that further work on this model
will help us to understand how a finite number of
inelastic process in a spatially separated region
compete with elastic processes (like interference
and resonance) and
affect persistent currents that show similar scaling
behavior as transport currents.

We have therefore shown that Fig.~1(b)  gives us a situation
where we can easily switch off or on the anti-resonances
and also strongly or weakly couple a reservoir
and thus study the exact effects produced by anti-resonances.
The presence or absence of anti-resonances are however
an universal feature of finite width quantum wires.

In summary, we have shown that anti-resonances can
drastically limit the broadening of eigen energies by an
electron reservoir and as a result, when the stub length
is made large the system bears no resemblance to a
ring-reservoir system. In absence of the anti-resonances,
the system can continuously go over to a ring-reservoir system
as the stub length is made large.
Although this is strictly valid when there
is a complete spatial separation of inelastic and elastic
processes, the two situation demonstrated in Fig.~3 and 6 are so
dramatically different that some of this effect will
survive even in presence of a finite number of inelastic
processes in the stub.  This effect may be relevant in the
dephasing effects observed in the experiment of Ref. \cite{sch}
where the zeroes were also observed.  We have also shown that
the anti-resonances has an anchoring effect on the poles
that can modify the distribution of poles in situations similar
to that considered in Ref. \cite{akk2}.


   
      \centerline{\Large {\bf Figure captions} }
      \ \\
\noindent Fig.~1 Three ways of attaching a reservoir to a system
of a ring coupled to a stub.

\noindent Fig.~2 Persistent current versus incident energy
at two different fields in a ring coupled to a 10 times longer
stub. The reservoir is attached close to the ring-stub junction. 

\noindent Fig.~3 Persistent current versus incident energy
in a ring coupled to a 100 times longer stub (dotted curve). 
The reservoir is attached close to the ring-stub junction. Thick
solid curve gives persistent current in the ring if it were
attached to an infinite reservoir.

\noindent Fig.~4 Persistent current versus incident energy
at two different fields in a ring coupled to a 10 times longer
stub (thick and thin solid curves).  The reservoir is attached
weakly close to the dead end of the stub. Thick and thin solid
curves show oscillations about dotted and dashed curves,
respectively, which are persistent currents in the same ring
coupled to an infinite reservoir.

\noindent Fig.~5 Same as in Fig.~4 but the reservoir is coupled
strongly.

\noindent Fig.~6 Same as in Fig.~5 but stub length is made 100
times the ring length.
   

\begin{thebibliography}{99}
\bibitem[*]{eml}email: deo@phys.jyu.fi   
\bibitem{par} T. P. Pareek and A. M. Jayannavar, 
Phys. Rev. B {\bf 57}, 8809 (1998).
\bibitem{lan} R. Landauer, Z. Phys. B {\bf 68} 217 (1987).
\bibitem{but1} M. B{\"u}ttiker, Phys. Rev. Lett. {\bf 57}, 1761 (1986).
\bibitem{but2} M. B$\ddot u$ttiker, Phys. Rev. B {\bf 32}, 1846 (1985).
\bibitem{mai} D. Mailly et al, Phys. Rev. Lett. {\bf 70}, 2020 (1993).
\bibitem{jay1} A. M. Jayannavar and P. Singha Deo, Phys. Rev. B
{\bf 50}, 13685 (1994).
\bibitem{jay2} A. M. Jayannavar and P. Singha Deo, Phys. Rev. B 
{\bf 51}, 10175 (1995); T. P. Pareek, P. Singha Deo 
and A. M. Jayannavar, Phys. Rev. B {\bf 52}, 
14657 (1995); S. Y. Cho and C. M. Ryu, 
Int. Journ. of Mod. Phys. B, {\bf 10} 3569 (1996).
\bibitem{mos} M.V.Moskalets, Europhys.Lett. {\bf 41}, 189 (1998).
\bibitem{deo1} P.Singha Deo, Phys. Rev. B {\bf 51}, 5441 (1995).
\bibitem{but3} M. B{\"u}ttiker and C. A. Stafford, Phys. Rev. Lett.
{\bf 76}, 495 (1996).
\bibitem{and} E. V. Anda, G. Chiappe and Ferrari, 
J. Phys.: Condens. Matter {\bf 9}, 1095 (1997).
\bibitem{leg} A.J. Leggett in: Granular nano-electronics,
eds. D. K. Ferry, J.R. Barker and C. Jacobony, NATO ASI Ser.
B {\bf 251} (Plenum, New York, 1991) p. 297.
\bibitem{deo3} P.Singha Deo, Solid St. Communication {\bf 107}, 
69 (1998); P. F. Bagwell, Phys. Rev. B, {\bf 41}, 10354 (1990);
E. Tekman and P. Bagwell, Phys. Rev. B {\bf 48} 2553 (1993).
\bibitem{gri} S. Griffith, Trans. Faraday. Soc. {\bf 49}, 650 (1953).
\bibitem{akk1} E. Akkermans, A. Auerbach, J. E. Avron
and B. Shapiro, Phys. Rev. Lett. {\bf 66}, 76 (1991).
\bibitem{deo2} P. Singha Deo, Phys. Rev. B {\bf 53}, 15447 (1996).
\bibitem{akk2} E. Akkermans and G. Montambaux, Phys. Rev. Lett.
{\bf 68}, 642 (1992).
\bibitem{bay} B. F. Bayman and C. J. Mehoke, Am. Journ. of Phys.
{\bf 51} 875(1983); W. Porod, Z. Shao and C. S. Lent, Phys. Rev. B
{\bf 48}, 8495(1993) and references therein.
\bibitem{kon} J. K{\"o}nig, Y. Gefen and G. Sch{\"o}n, Phys.
Rev. Lett. {\bf 81}, 4468 (1998).
\bibitem{pri} A. M. Jayannavar, private communication (1994).
\bibitem{sch} R. Schuster, E. Buks, M. Heiblum, D. Mahalu
and V. Umansky, Nature {\bf 385}, 417 (1997);
E. Buks, R. Schuster, M. Heiblum, D. Mahalu and V. Umansky,
Nature {\bf 391}, 871 (1998).


    \end{thebibliography}
      
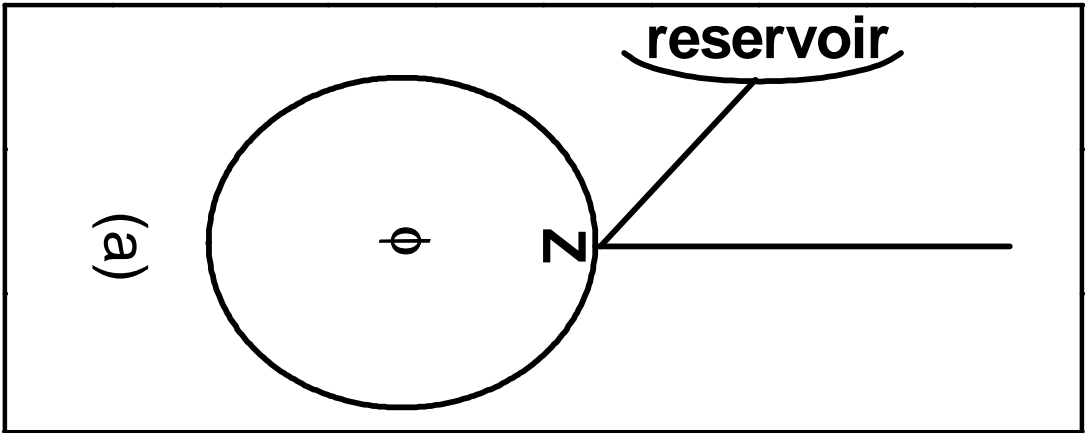
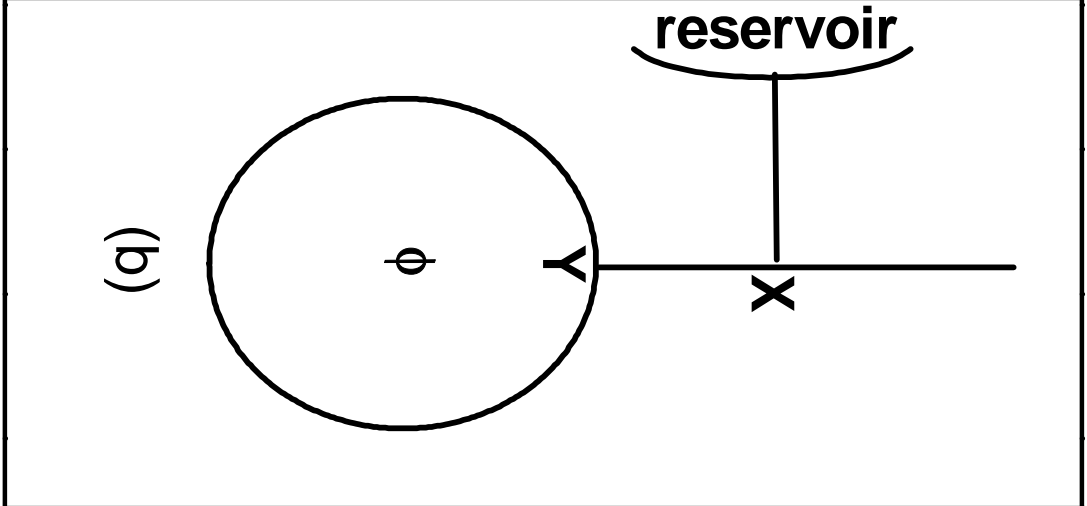
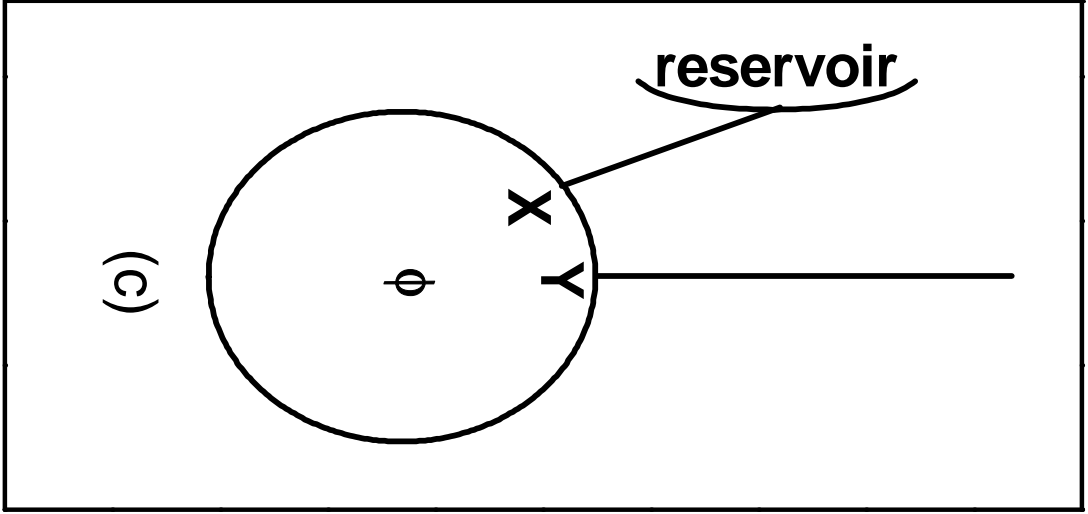
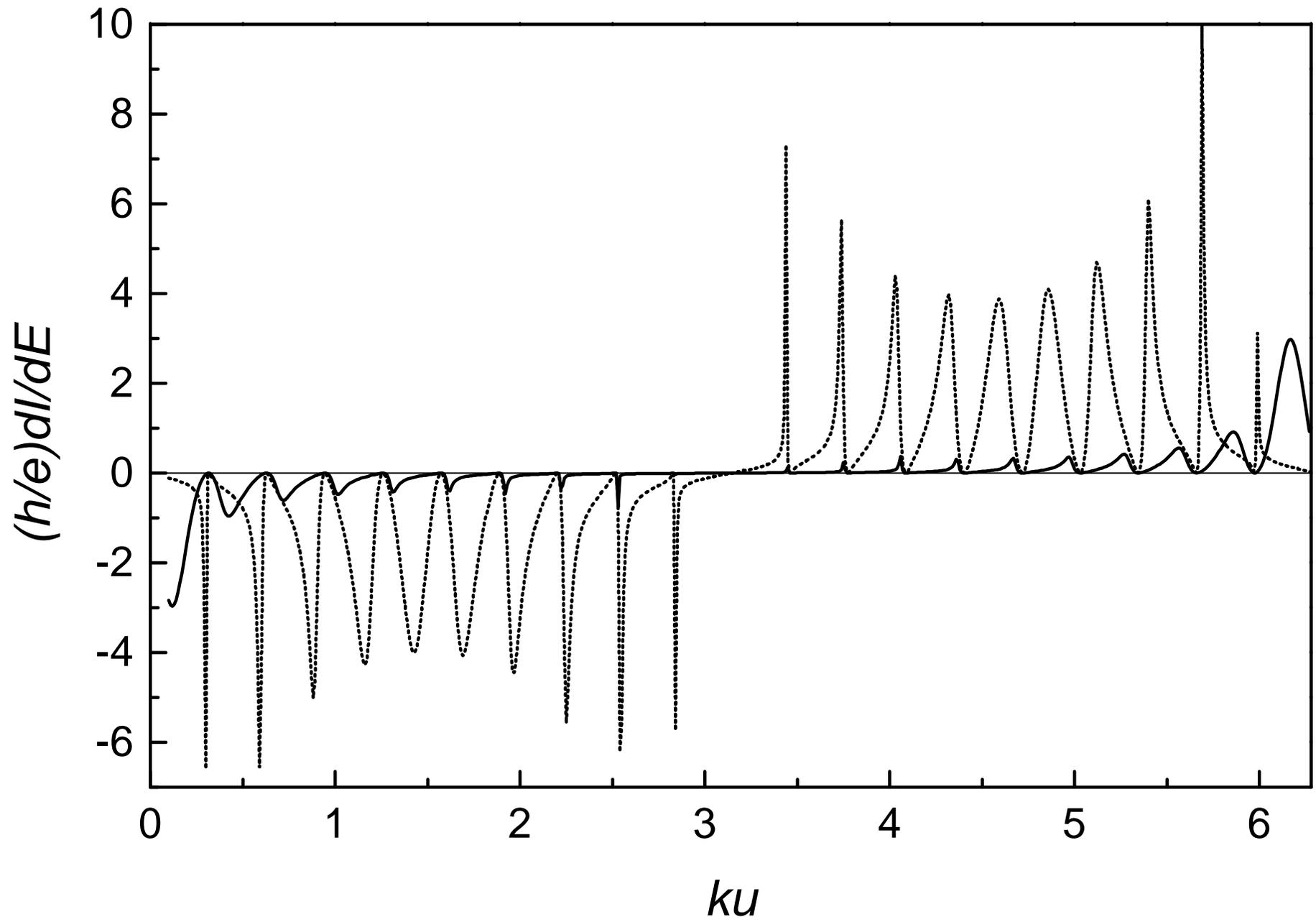
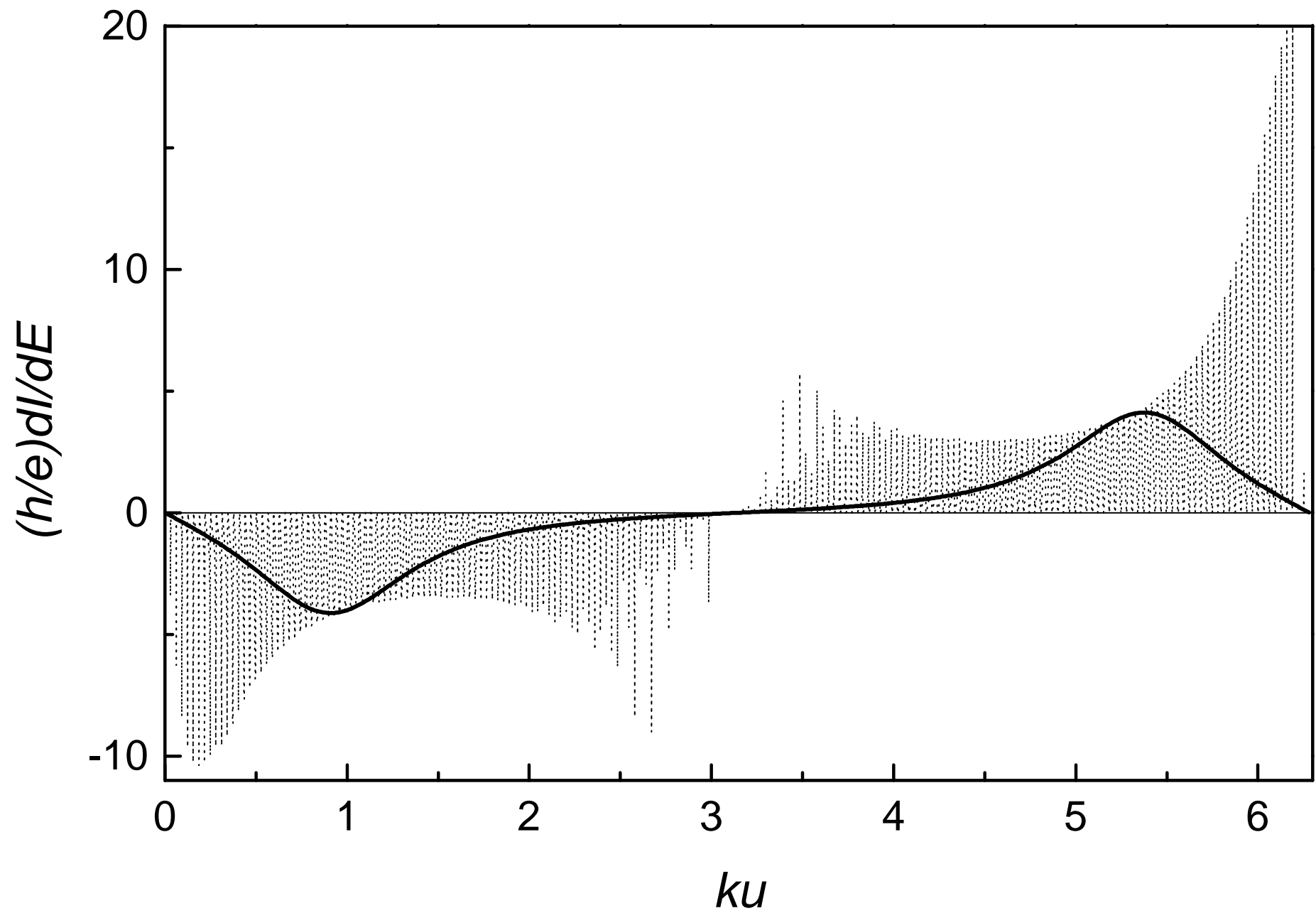
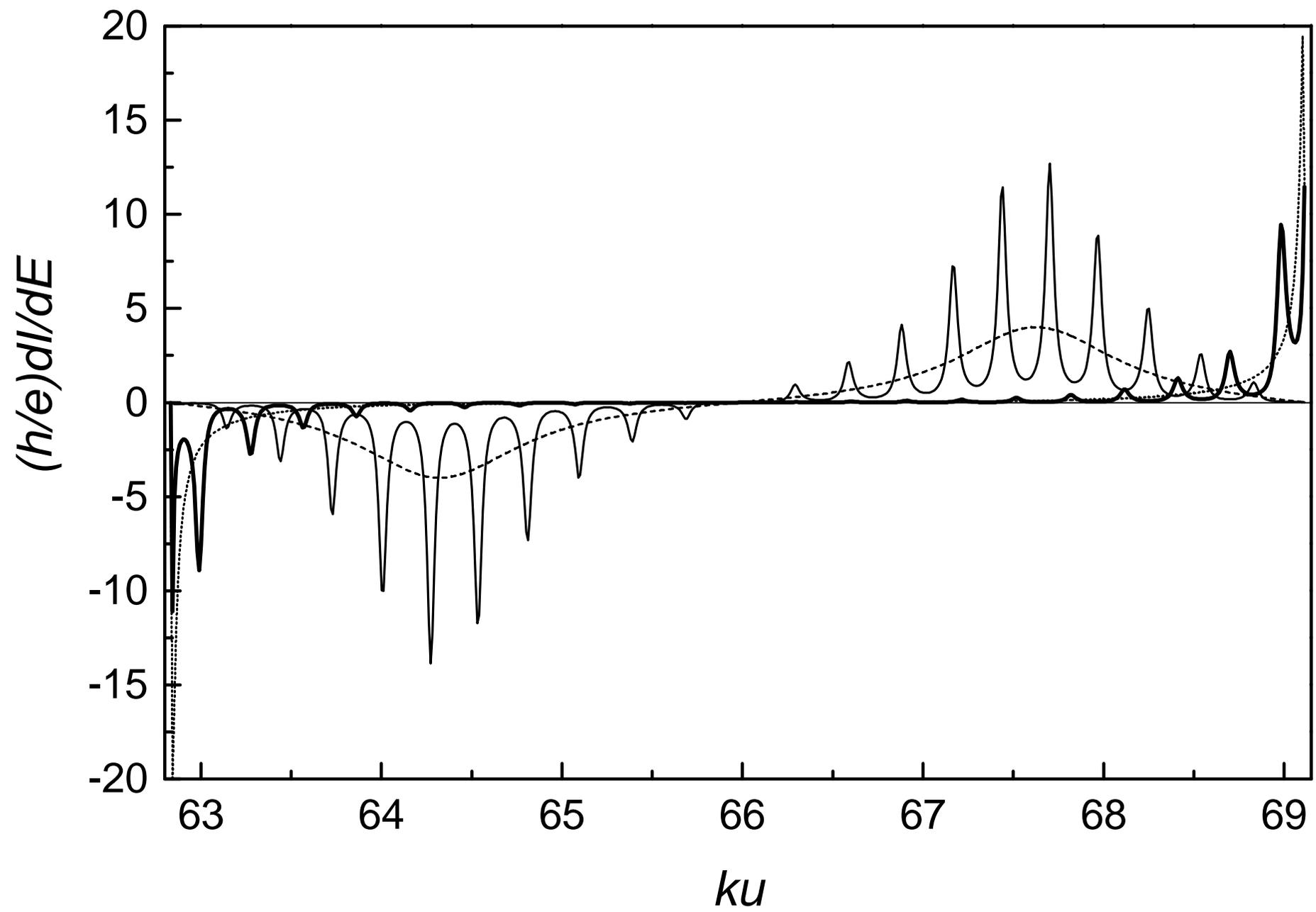
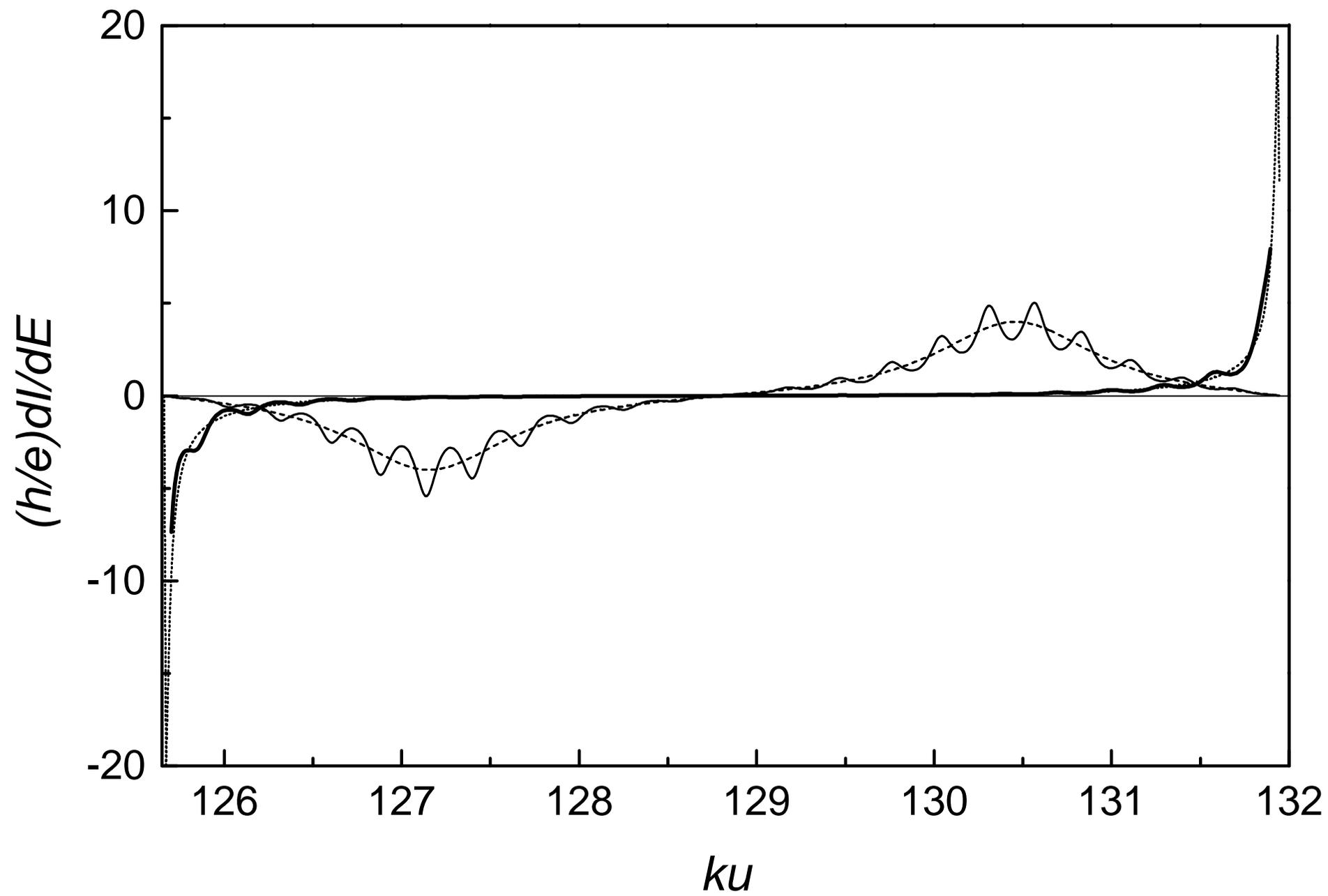
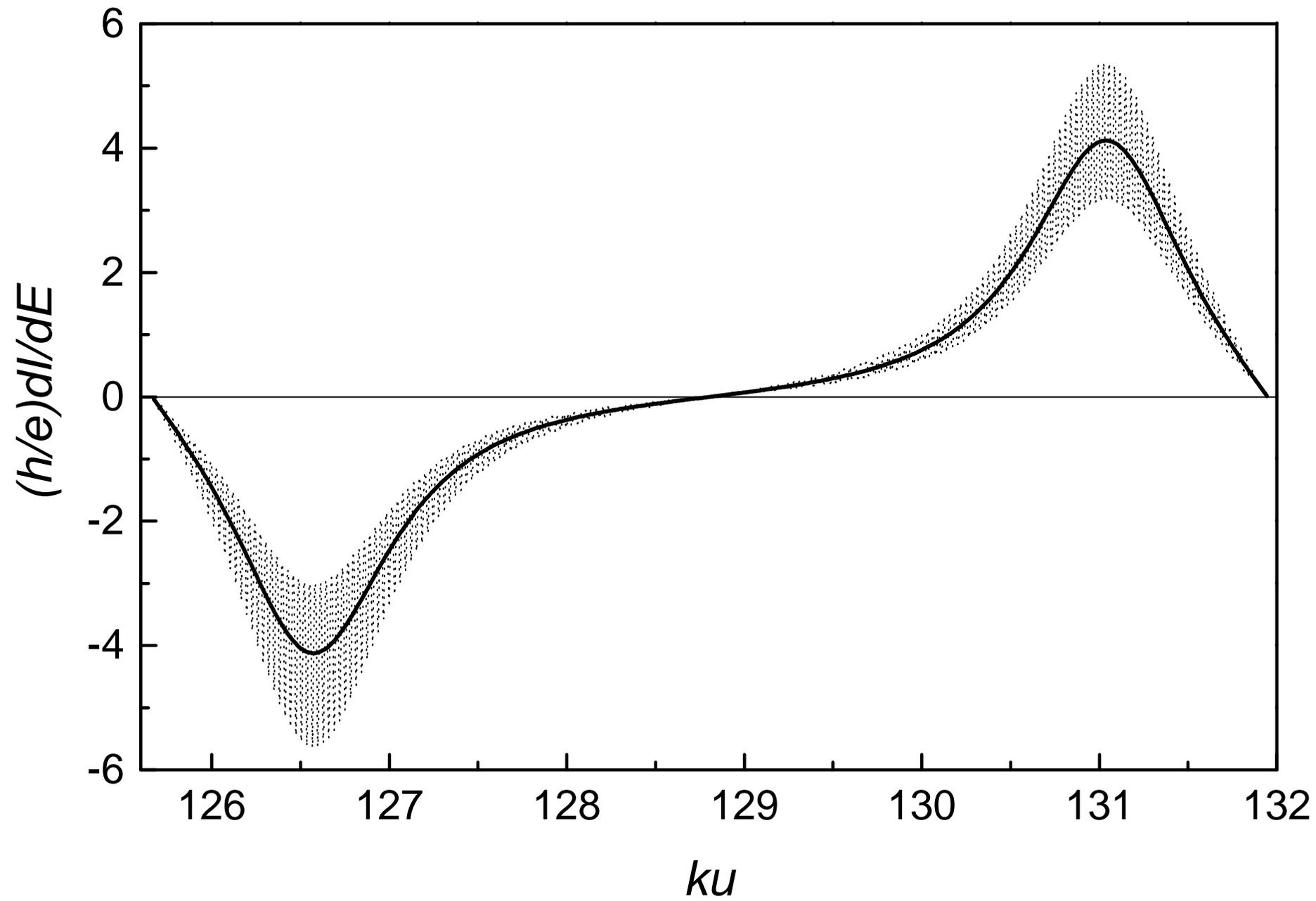
\end{document}